\documentclass[
aps,
12pt,
final, 
notitlepage,
oneside,
nobibnotes,
nofootinbib,
superscriptaddress,
noshowpacs,
centertags,
tightenlines% for normal intervals
]
{revtex4}
 
\newcommand{\aap}{A\&A}

\newcommand{\aaps}{A\&AS}

\newcommand{\aj}{AJ}
\newcommand{\apjs}{ApJS}

\newcommand{\mnras}{MNRAS}

\usepackage[T2A]{fontenc}
\usepackage[utf8]{inputenc}
\usepackage{amsmath}
\usepackage{subfig}
\usepackage{wrapfig}
\usepackage{gensymb}
\usepackage{graphicx}

\begin{document}

\title{Study of Star-Forming Regions in the Peculiar Galaxies NGC 660, NGC 1512, NGC 4395, and NGC 4618}

\author{\firstname{K.~I.}~\surname{Smirnova}}
\email{Arashu@rambler.ru}
\affiliation{Ural Federal University, Yekaterinburg, Russia}

\author{\firstname{D.~S.}~\surname{Wiebe}}
\affiliation{Institute of Astronomy, Russian Academy of Sciences, Moscow, Russia}

\author{\firstname{A.~V.}~\surname{Moiseev}}
\affiliation{Special Astrophysical Observatory, Russian Academy of Sciences, Nizhnii Arkhyz, Russia}

\author{\firstname{G.~I.~G}~\surname{J\'ozsa}}
\affiliation{South African Radio Astronomy Observatory, Cape Town, South Africa}
\affiliation{Department of Physics and Electronics, Rhodes University, Makhanda, South Africa}
\affiliation{Argelander-Institut für Astronomie, Bonn, Germany}
\begin{abstract}
The star-forming regions (SFRs) in the peculiar galaxies NGC 660, NGC 1512, NGC 4395 and
NGC 4618 are studied. The relationships between such characteristics of star forming regions as UV, nearand far-IR fluxes, as well as in the H$\alpha$ and HI lines, surface brightness in these ranges, and the scatter of the radial velocities of ionized and neutral hydrogen are considered. It is shown that in all the galaxies considered, the IR fluxes from SFRs are less than in the “normal” galaxies, but for different reasons: in the galaxies with signs of recent interaction NGC 660 and NGC 1512, this is due to the low surface brightness of SFRs; in the lopsided galaxies NGC 4395 and NGC 4618, the low brightness of SFRs in the infrared range is due to their compact size. These differences indicate that the star formation process depends not only on the morphological type of a galaxy, but also on many other factors.

\end{abstract}

\maketitle

\section{INTRODUCTION}

Modern space and ground-based instruments allow spatially resolved observations of interstellar matter in extragalactic star-forming regions (SFR), which opens up opportunities for studying the features of the star formation process in a significantly wider range of parameters than is possible in our Galaxy. In particular, it is possible to study star-forming in detail in those galaxies in which the birth of stars is significantly affected by some external influence or even stimulated by such influence, for example, in interacting systems. 

Observations show that the characteristics of star formation in galaxies of different morphological types differ, and interstellar matter in them is distributed differently: the star-forming regions can be located both chaotically and ordered. Of particular interest from this point of view are those galaxies that show signs of recent triggered star formation, for example, as a result of tidal interaction or merging with another galaxy. Such objects may include so-called polar-ring galaxies—a rather rare class of objects whose subsystems (disk and ring) rotate in almost orthogonal planes. In some close galaxies of this type it is possible to trace differences between star-forming regions and stellar clusters in the ring and in the disk \cite{2004A&A...421..833K,2016A&A...585A.156O}. Although the reasons for the formation of polar subsystems may be different, and there are ongoing disputes about the nature of polar rings in specific galaxies  \cite{BournaudCombes2003,2019MNRASEgorovMoiseev}, in any
case, we are talking about the fact that a galaxy captures external matter with a moment of rotation, which is very different from that of the disk.

In \cite{Smirnova2017,2019ARep...63..445S} we considered various relationships between the observed characteristics of “normal ” galaxies. The subject of this study is similar relations for those galaxies, in the history of which there were episodes that led to certain features in their modern morphology and kinematics. This study was started by us in \cite{2017SWM}, devoted to the study of star formation processes in the ring and disk of the polar-ring galaxy NGC 660. We have revealed some patterns in the emission parameters of the ring and disk SFRs in the near- and mid-IR ranges. In particular, it turned out that the infrared flux from SFRs in the ring and disk of the galaxy differ by orders of magnitude.

Unfortunately, we were not able to find at least one more polar-ring galaxy to continue this analysis, since it requires that the ring be sufficiently extended to allow the detection of as many SFRs as possible, which would not blend with each other when observed by the Spitzer and Herschel Space IR Observatories with an angular resolution of $2-12''$.

In other words, it is necessary that in a galaxy it is possible to fulfill photometry of regions corresponding in size to individual SFRs, or at least to a complex of close SFRs. In
addition, there should be sufficient observational data in other ranges for the selected galaxy. To date, there are several hundred known polar-ring galaxies, but they are all too far away. Moreover, only for a part of the candidates selected by their peculiar morphology, there are reliable kinematic confirmations of the presence of polar rings \cite{1990Whitmore, 2011Moiseev, 2019MNRASReshetnikov}. 

So we had the idea to supplement the analysis with other types of galaxies. In the work by \cite{2017SWM} it was concluded that the differences in the fluxes from the disk and ring SFRs of NGC 660 are related to differences in their ages, namely: the regions of the ring in which the near-and mid-IR fluxes are 1‒2 orders of magnitude less than in the disk SFRs of NGC 660 (as well as in the SFRs of “normal” highmetal galaxies) are younger. Based on this assumption, we looked for galaxies that had dynamic processes in the recent past that could cause a wave of star formation. Such galaxies should be located at a short distance (no more than 15 Mpc), so that individual SFRs can be resolved in them, and it is desirable to have a “face on” orientation, in order to simplify the allocation of SFRs and reduce the line-of-sight confusion. Based on the above criteria, we included in consideration the interacting galaxy NGC 1512, for which data is available not only in the near-infrared, but also in the far-infrared ranges.

In addition, we paid attention to asymmetric (lopsided) galaxies \cite{2009PhR...471...75J}. Numerical simulations show that asymmetry can occur when a galaxy interacts with a companion \cite{2014MNRAS.439.1948Y} or as a result of the interaction of two dwarf galaxies \cite{2016Pardy}, resulting in the galaxy’s disk becoming “lopsided”. Disk asymmetry in this model persists for almost 2 billion years. One type of lopsided galaxy is a galaxy with an offset bar \cite{2017MNRAS.469.3363K}, in which the position of the bar does not coincide with the photometric center of the galaxy disk. Unfortunately, among galaxies with an offset bar only the galaxy NGC 4618 meets our criteria \cite{1991AJ....101..829O}. Another asymmetric galaxy with the necessary data set was the Seifert galaxy NGC 4395 \cite{1995ApJS...98..477H}. Both of these galaxies belong to the SBm type.

In the previous work \cite{Smirnova2017} we investigated the relationship between the various components of the interstellar medium (ISM) (atomic and molecular hydrogen, dust particles of various types), which are directly involved in the star formation process, in eleven galaxies, which were divided into two groups: with high and low metallicity (metallicity was taken from the article \cite{moustakas2010}). In this paper, for comparison with our previous results, we consider the high-metallicity galaxy NGC 628. It is further called the comparison galaxy, while the galaxies NGC 660, NGC1512, NGC 4395 and NGC 4618 are collectively called new galaxies.

To compare the surface brightness of SFRs in different galaxies, it is necessary to know the distances to them. For uniformity, we have accepted estimates from the Cosmicflows-3 catalog \footnote{http://edd.ifa.hawaii.edu} \cite{2016AJ....152...50T}. They are shown in Table~\ref{distances} along with the corresponding angular scales. The table also provides estimates of the stellar mass in the galaxies under consideration, derived from fluxes at wavelengths of 3.6 and 4.5 $\mu$m (the Spitzer telescope) using the method described in the paper \cite{2012AJ....143..139E}. Since for our study it is sufficient to know the approximate mass values (see Section 4), we did not attempt to perform accurate photometry. However, our estimates are close to the values obtained for NGC 628 and NGC 660 in  \cite{2019PASJ...71S..14S} and for NGC 4395 and NGC4618 in \cite{2016MNRAS.463.2092Y}. In addition, Table~\ref{distances} gives an estimate of the total star formation rate derived from the flux at a wavelength of 24 $\mu$m according to the calibration given by \cite{2013seg..book..419C}, for all galaxies except NGC 660, for which data at a wavelength of 24 $\mu$m  are not available.

The metallicity of the galaxies NGC 660 and NGC 4618 was calculated using calibration ratios for determination of the oxygen content from R-calibrations from the work \cite{2016Pilyugin}
\begin{eqnarray}
R_2 = I_{{\rm O}_{\rm II}} \lambda 3727+\lambda 3727/I_{{\rm H}_{\beta}}, \nonumber\\
N_2 = I_{{\rm N}_{\rm II}} \lambda 6548+\lambda 6584/I_{{\rm H}_{\beta}}, \nonumber\\
S_2 = I_{{\rm S}_{\rm II}} \lambda 6717+\lambda 6731/I_{{\rm H}_{\beta}}, \nonumber\\
R_3 = I_{{\rm O}_{\rm III}} \lambda 4959+\lambda 5007/I_{{\rm H}_{\beta}}.
\label{main_difeq}
\end{eqnarray}
The line intensities for these galaxies were taken in the article \cite{2006Moustakas}.

The last column of Table~\ref{distances} indicates whether the galaxies in question belong to groups or clusters. The opinions of different authors on the attribution of the galaxies NGC 660 and NGC 628 to one or to different groups differ. However, even if both of these galaxies belong to the same group, they are spatially far apart, and the effects of the environment should not be large.

\begin{table*}
%\begin{center}
\caption{Characteristics of the studied galaxies}
\label{distances}
\begin{tabular}{lccccccc}
\hline
Galaxy & Distance, & Angular & 12 + & Morph. & $\lg(M_*/M_\odot)$ & Rate &Group,\\
 & Мпк & scale, &  lg(O/H) &type &  &  of SF, &cluster\\
 &   & pc/arcsec  &  &  &  & $M_{\odot}$/year &\\
\hline
NGC 660  & 13.55 & 66   & 8.89  & SB(s)a pec & 10.5 & & LGG29 \cite{1993Garcia},\\
         &       &      &       &             &   & & NGC 660 \cite{2011Makarov}\\
\hline
NGC 1512 & 12.25 & 60   & 8.56 \cite{moustakas2010}  & SB(r)a; HII & 10.2 & 0.4 & LGG108 \cite{1993Garcia},\\
         &       &      &       &             &   & &  NGC 1512 \cite{2011Makarov}\\
\hline
NGC 4395 & 4.76  & 23   & 8.33 \cite{1996Roy} & SA(s)m; & 9.4 & 0.03 & LGG291 \cite{1993Garcia},\\
         &       &      &       & LINER Sy1.8  & & & USGCU480 \\
\hline
NGC 4618 & 6.52  & 32  & 8.35 & SB(rs)m; HII & 9.7 & 0.05 & LGG290 \cite{1993Garcia}\\
\hline
NGC 628  & 9.77  & 47  & 8.45 \cite{moustakas2010} & SA(s)c; HII  & 10.4 & 1.0 & LGG29 \cite{1993Garcia},\\
         &       &      &       &             &   & & NGC 628 \cite{2011Makarov}\\
\hline
\end{tabular}
%\end{center}
\end{table*}

\section{OBSERVATIONAL DATA AND REDUCTION}

We used the same sets of observations of the NGC 660 galaxy in the near- and mid-IR ranges, as well as
in the H$\alpha$ line as \cite{2017SWM}. In this article, in addition to the integral intensity of the line H$\alpha$ we also consider the spectra of this emission to estimate the velocity dispersion of the ionized gas. The observations were performed on December 22–23, 2003, at the SAO RAS 6-meter telescope using a Fabry–Perot scanning interferometer in the multimode focal reducer SCORPIO \cite{Afanasiev2005}. The width of the interfringe was about $1290 km/s$ around the red-shifted emission line H$\alpha$. This range was evenly filled with 32 interferograms with exposures of 240 s each. The result of processing performed using the software package \cite{MoiseevEgorov2008}, is a data cube where each pixel of size 0.71" in the field of view $6.1'\times6.1'$ contains a 32-channel spectrum. The spectral resolution was $120 km/s$, and the angular resolution was 2.9".

\begin{figure}[t!]
    \begin{subfloat}
    \centering{\includegraphics[width=0.8\linewidth]{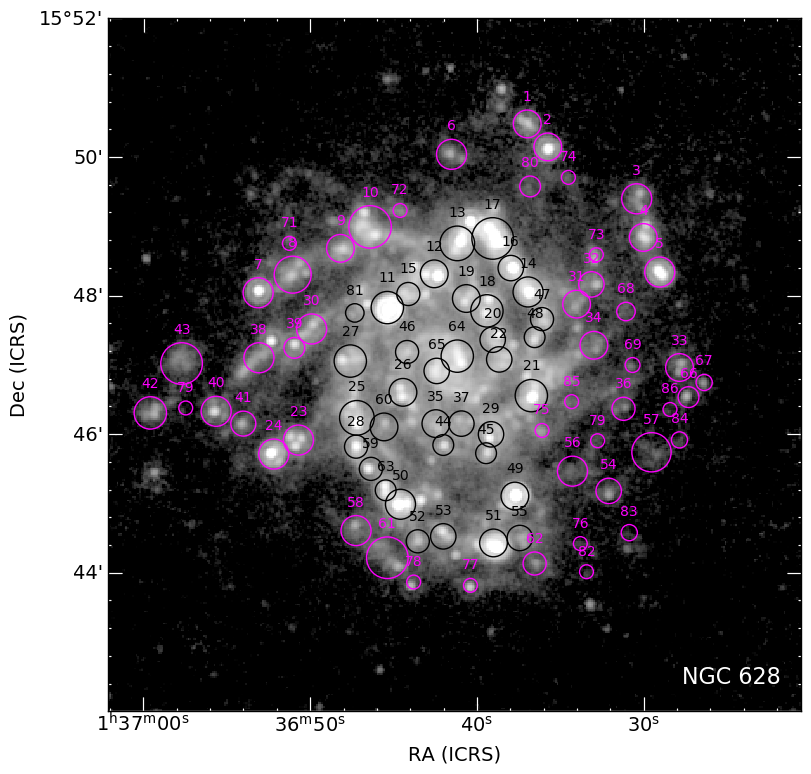}} 
    \end{subfloat}
\caption{Star-forming regions of the galaxy NGC 628: the black circles are regions from the central part of the galaxy, the purple circles are peripheral regions. An image of the galaxy at a wavelength of 24 $\mu$m from the Spitzer telescope’s science archive is used as the background.}
\label{NGC628_dopreg}
\end{figure}

Data sources for the galaxy NGC628 are given in \cite{Smirnova2017,2019ARep...63..445S}. In them, we imposed a limitation on the size of the studied areas (at least 12) due to the use of observational
data in the CO line. As a result, we selected 65 SFRs bright in IR in the galaxy NGC 628. In this paper, we do not consider the CO radiation, and, in addition, for comparison with the systems under consideration, we need more faint ones in the IR band. Therefore, we additionally identified 19 more regions in the galaxy NGC 628. The entire set of regions under consideration is shown in Fig.~\ref{NGC628_dopreg}. Different colors show the areas of the central part of the galaxy and the areas on its periphery (see below).

Near-IR observations for the galaxies NGC1512, NGC 4395 and NGC 4618 (at wavelengths of 3.6, 4.5, 8 and 24 $\mu$m) were taken from the Spitzer telescope’s scientific archive. For the galaxy NGC 660, Spitzer observations at a wavelength of 24 $\mu$m are not available, so we used data obtained using the WISE telescope instead. Data from observations of all galaxies in the ultraviolet range were downloaded from the GALEX telescope archive. For the galaxy NGC 1512, data are also available in the far IR range from the KINGFISH survey\footnote{http://herschel.esac.esa.int/Science\_Archive.shtml} \cite{kingfish}, performed on the Herschel telescope.

The observational data used had to be reduced (if possible) to a single angular resolution. In the near and middle IR ranges, the worst angular resolution was obtained from observations of NGC 660 with the  WISE telescope at a wavelength of 22 $\mu$m, so observations of all galaxies at wavelengths of 8 and 24 $\mu$m, as well as observations of the galaxy NGC1512 at wavelengths of 70 and 100 $\mu$m, were reduced to the resolution in the photometric band at a wavelength of 22$\mu$m ($\sim12''$). The convolution was made using kernels from the work of \cite{2011Aniano}. Observations of the galaxy NGC 1512 at a wavelength of 160 $\mu$m were used in their original form, since their angular resolution is close to the angular resolution of the data at a wavelength of 22 $\mu$m.

We also have data in the HI line for the galaxy NGC 660. It was observed in the L band using the Westerbork Synthesis Radio Telescope (WSRT) for three epochs in December 2009, 12 hours for each epoch. The total bandwidth was 10 MHz and was divided into 1024 channels. In accordance with standard practice for WSRT, the primary calibrators CTD93 and 3C147 were observed for approximately 30 minutes before and after the main object was observed. Standard data processing was performed using the software {\sc Miriad} \cite{1995ASPC...77..433S}. After performing initial cross-calibration of the gain and bandwidth using the observed calibration sources, we used an iterative self-calibration procedure to improve the gain calibration in the continuum: first, the image was constructed in the continuum using homogeneous data weighting, which was used for subsequent iterative cleaning by the method of CLEAN components \cite{1974A&AS...15..417H,1980A&A....89..377C}. After each CLEAN procedure, we defined CLEAN areas by applying a cut-off at some level to the resulting image, which allowed us to allocate CLEAN components only within these areas. Using the resulting set of CLEAN components as the input model, we applied a self-calibration procedure to the data and repeated the cleaning of the image by the CLEAN method again. The process was repeated until the desired image quality was achieved (with the cut-off level gradually decreasing). The obtained values of complex gain coefficients in the continuum were then transferred to the original data set, and the resulting model was subtracted in the UV region. After additional subtraction of the continuum {\sc uvlin }in the UV region, the Hanning smoothing was applied to the data. The data cube was created using robust weighing \cite{briggs} and tapering. Due to the small declination of NGC 660, the projections of the WSRT bases in the East-West direction were greatly shortened, so it was necessary either to make the synthesized directional diagram symmetrical, thereby worsening the angular resolution (but losing in sensitivity significantly less than for sources with large declinations), or to leave it strongly elongated.
The data cube analyzed here had an RMS noise of
$\sigma_{\rm rms}\,=0.68\,\mathrm{mJy}\,\mathrm{beam}^{-1}$ and a resolution of $64^{\prime\prime}\times 12^{\prime\prime}$ (HPBW). Observations of neutral hydrogen were used without convolution and without taking into account the elongation of the beam. Our preliminary calculations showed that this does not change our conclusions.

\begin{figure*}
    \begin{subfloat}
    \centering{\includegraphics[width=0.4\linewidth]{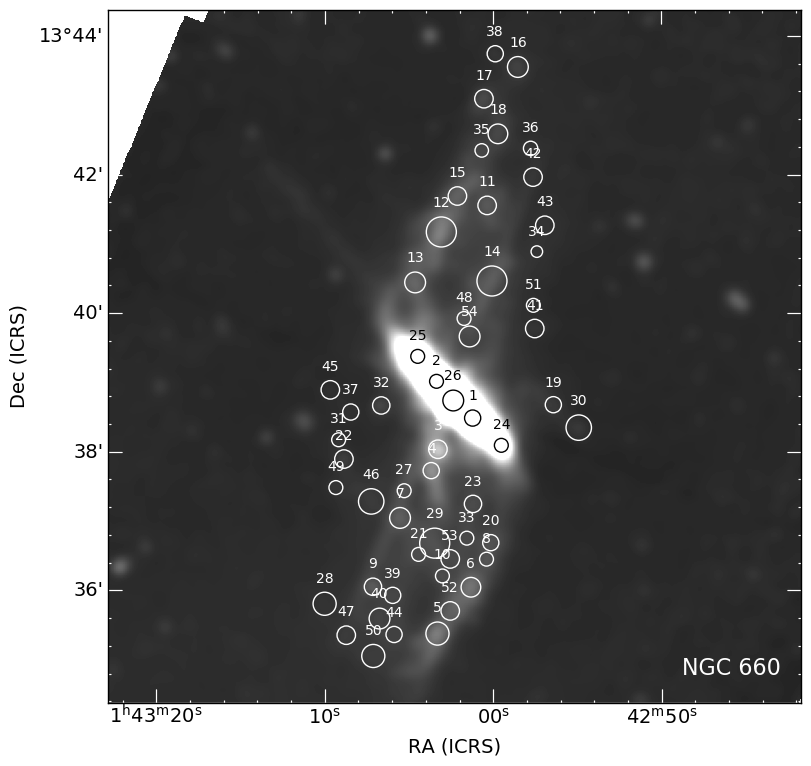}} 
    \end{subfloat}
    \begin{subfloat}
    \centering{\includegraphics[width=0.4\linewidth]{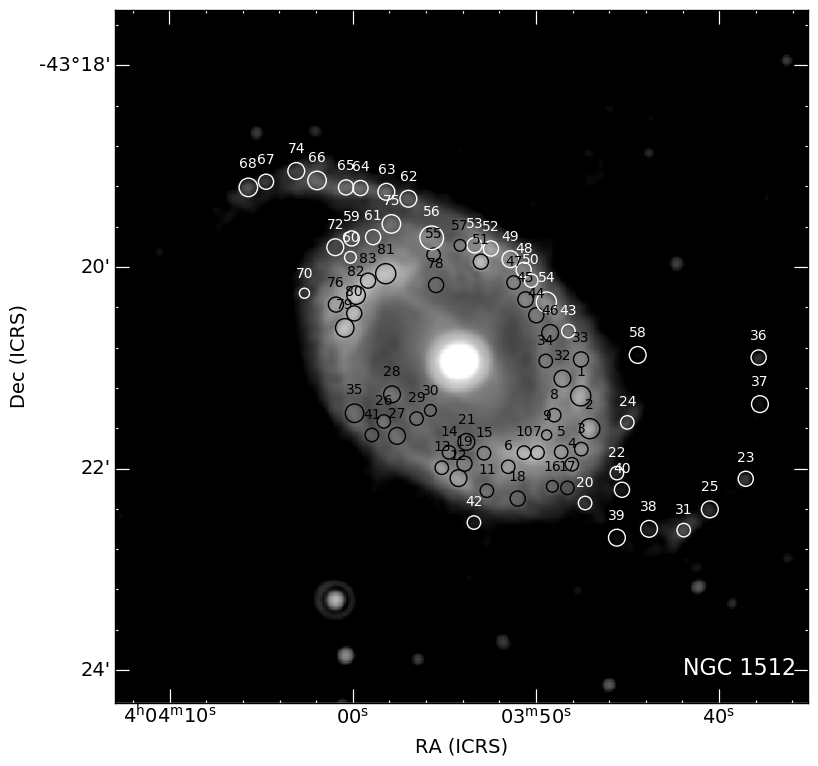}} 
    \end{subfloat}
    \begin{subfloat}
    \centering{\includegraphics[width=0.4\linewidth]{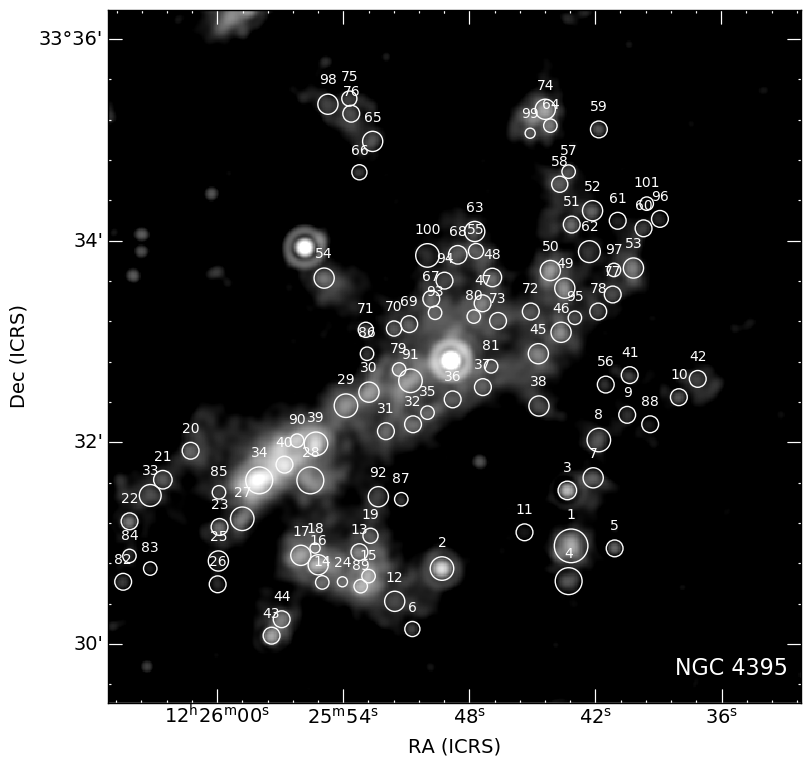}} 
    \end{subfloat}
    \begin{subfloat}
    \centering{\includegraphics[width=0.4\linewidth]{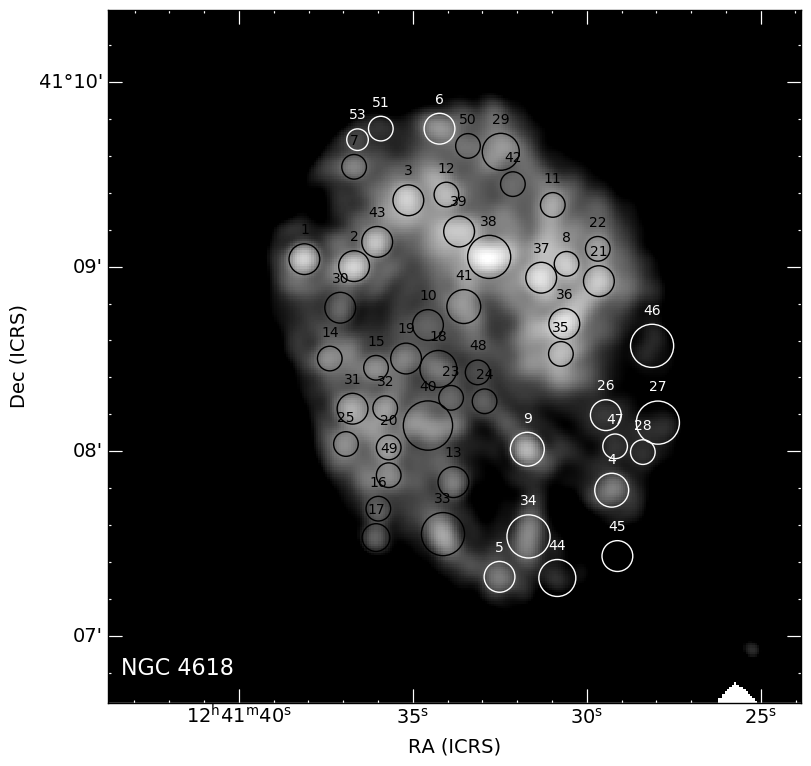}}
    \end{subfloat}
\caption{Star-forming regions in the galaxies considered in this paper, against the background of emission maps at a wavelength of 8 $\mu$m, convolved to the resolution of the WISE telescope at a wavelength of 22 $\mu$m.}
\label{regions}
\end{figure*}

The regions selected in the new galaxies are shown in Fig.~\ref{regions}. After preparing the observation material and selecting the studied regions, we performed aperture photometry with background subtraction. The photometry procedure and estimation of measurement errors are described in \cite{pahcycle}. The background in all cases was estimated by a six-pixelwide ring surrounding the study region. Given the pixel size for IR images, this means that the background detection ring is about twice as wide as the aperture itself. This wide ring allows you to smooth out the effect of neighboring bright areas and correctly take into account the diffuse background in the aperture.

As a result, radiation fluxes in the near and middle IR range, as well as in the UV range, were estimated for all galaxies. The fluxes in bands at wavelengths of 8 and 24 (22) $\mu$m were corrected for the contribution of stellar radiation. In \cite{pahcycle}, the notations $F_{8}^{\rm afe}$ and $F_{24}^{\rm ns}$ were used for corrected fluxes. Further, the upper indexes are omitted for simplicity, but the comparison is made with the corrected fluxes. For the galaxy NGC 1512, the fluxes at wavelengths of 70, 100, and 160 $\mu$m were also estimated.

\section{RESULTS}

\subsection{Aperture Photometry}

The main object of our research is the observed properties of dust and gas in the star-forming regions. We assume that the source of radiation in the near-IR range is small aromatic particles (polycyclic aromatic hydrocarbons, PAHs), whereas radiation at a wavelength of 24 (22) $\mu$m is generated by larger hot dust grains. In both cases, the generation of IR radiation requires heating the dust particle with ultraviolet radiation, so it can be assumed that the regions of intense radiation in the near and middle IR ranges are star-forming complexes.

\begin{figure}
    \begin{subfloat}
    \centering{\includegraphics[width=0.8\linewidth]{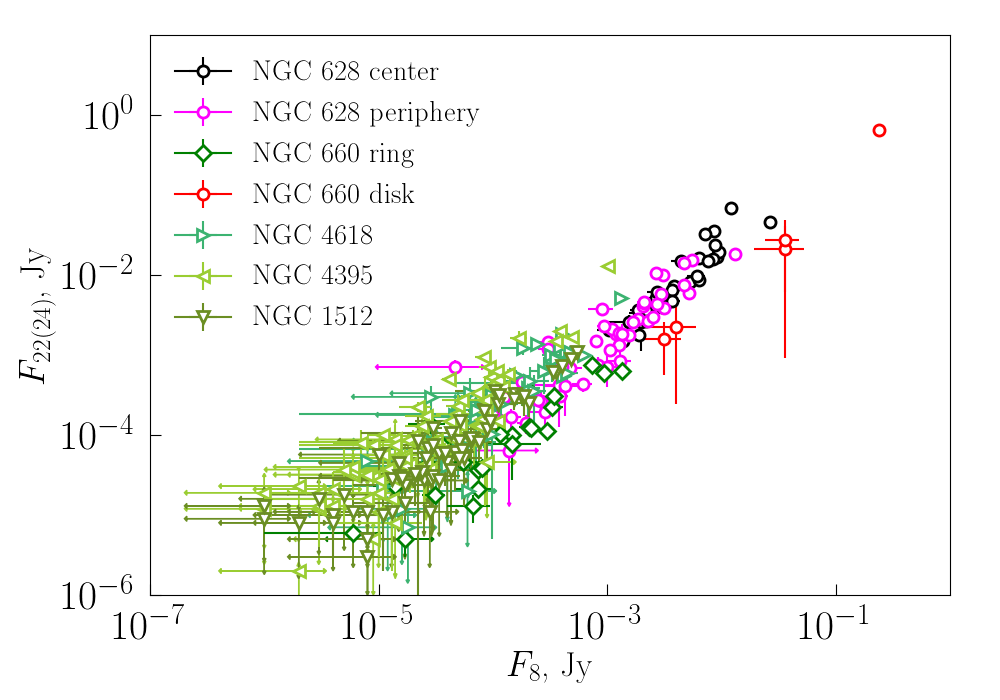}} 
    \end{subfloat}
\caption{The ratio of radiation fluxes at wavelengths of 8 and 24 (22) $\mu$m for SFRs in the considered galaxies NGC1512, NGC 4395, NGC 4618, NGC 660 (the ring and disk are marked with different symbols) and in the comparison galaxy NGC628.}
\label{F8_F22(24)}
\end{figure}

The common nature of emission at wavelengths of 8 and 24 $\mu$m is emphasized by the diagram in Fig.~\ref{F8_F22(24)}. It can be seen that the position of the SFRs in NGC 628 and the disk of the galaxy NGC 660 in this diagram differs from the position of the SFRs in the galaxies NGC 1512, NGC 4395, NGC 4618, as well as in the ring of the galaxy NGC 660. In the SFRs of new galaxies, the fluxes at wavelengths of 8 and 24 $\mu$m also correlate well with each other, but their values are significantly lower than the fluxes in the SFRs of the “ordinary” galaxy of high metallicity (especially its central part) and of the disk of the galaxy NGC 660.

\begin{figure*}
    \begin{subfloat}
    \centering{\includegraphics[width=0.4\linewidth]{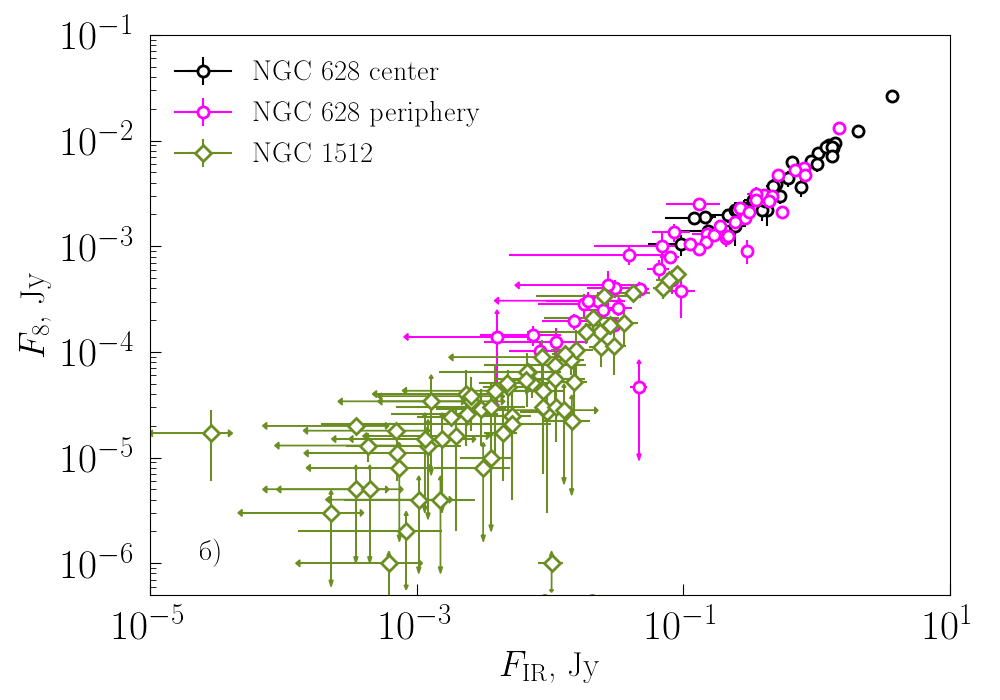}} 
    \end{subfloat}
    \begin{subfloat}
    \centering{\includegraphics[width=0.4\linewidth]{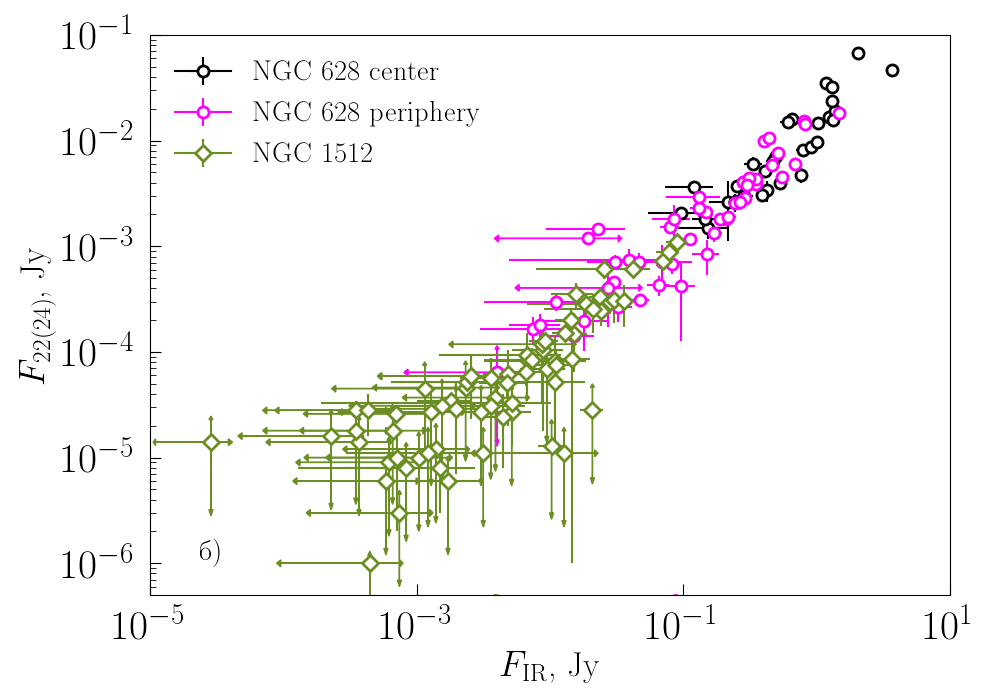}} 
    \end{subfloat}
\caption{The ratio of radiation fluxes at wavelengths of 8 and 24 $\mu$m and the total flux in the far IR range for SFRs in the galaxies NGC 1512 and NGC 628.}
\label{F8_24_FIR}
\end{figure*}

Since the Herschel data are also available for the NGC 1512 galaxy, we can see how its near and far IR range fluxes compare with “normal” galaxies. In Fig.~\ref{F8_24_FIR}, the total flux in the far IR range (the sum of fluxes at wavelengths of 70, 100, and 160 $\mu$m) is compared with fluxes at wavelengths of 8 and 24 $\mu$m. It is obvious that the SFRs in the galaxy NGC 1512 in the far IR range are also significantly dimmer than the SFRs in the comparison galaxy (both central and peripheral), although in comparison with the fluxes at wavelengths of 8 and 24 $\mu$m, they show the same trend.

\begin{figure}
    \begin{subfloat}
    \centering{\includegraphics[clip=,width=0.7\linewidth]{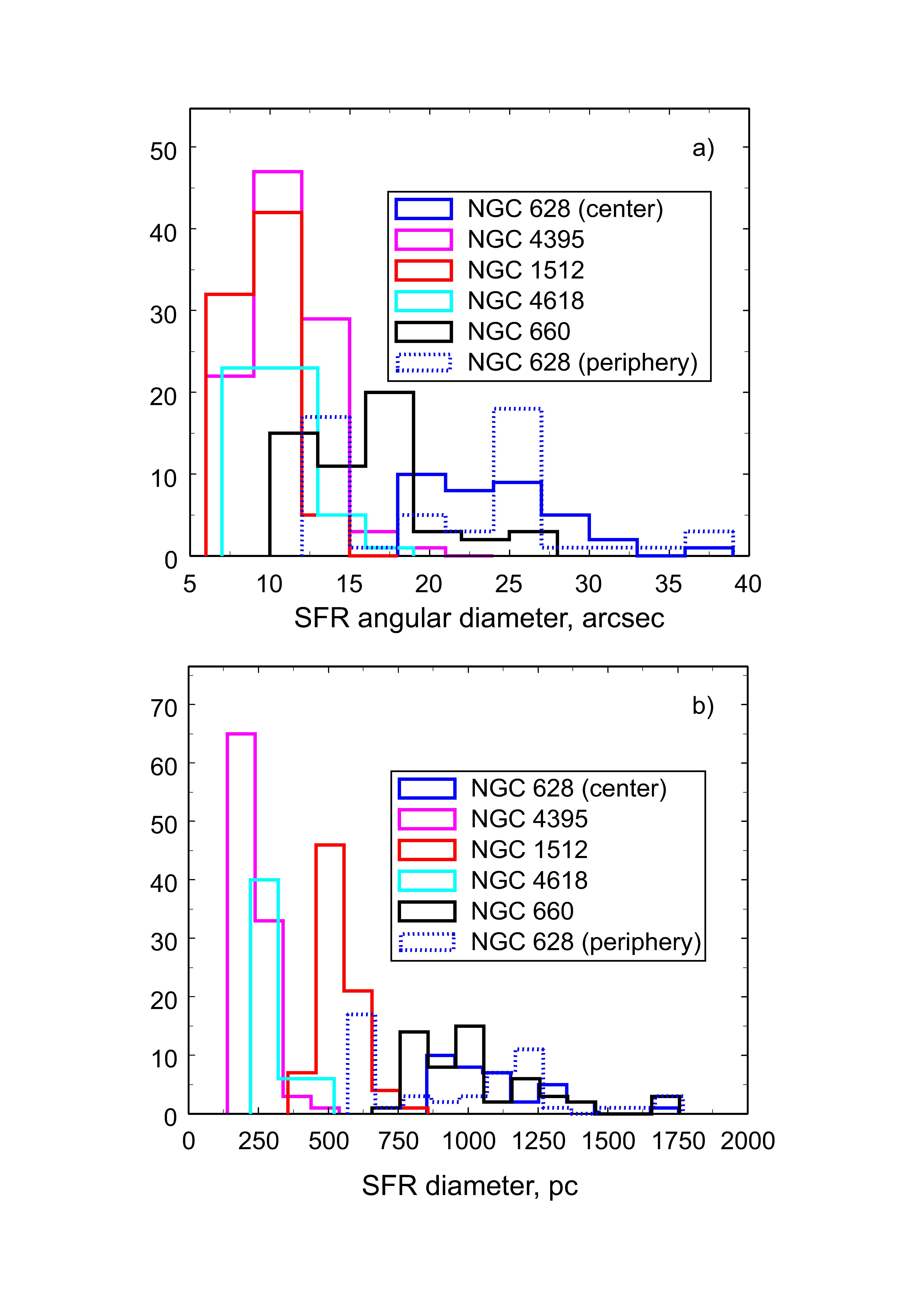}} 
    \end{subfloat}
\caption{Distributions of the studied SFRs by aperture diameters (a) and by linear diameters (b).}
\label{aperdiam}
\end{figure}

A direct interpretation of Figs.~\ref{F8_F22(24)} and \ref{F8_24_FIR} is that the infrared fluxes of SFRs in new galaxies are systematically lower than similar fluxes from SFRs in the comparison galaxy. However, this may be partly due to the fact that we have isolated SFRs of different sizes in different galaxies. It should be noted that in all cases this was done by eye, without using any formal algorithm (see the discussion in \cite{2017ASPC..510..102S}), however, we tried to select the size of the apertures so that they correspond to the size of the selected SFR as much as possible. Fig.~\ref{aperdiam}a shows histograms of aperture diameters for the galaxies in question. It is obvious that in the new galaxies, we systematically chose more compact SFRs than in the comparison galaxy. In NGC 628, a significant part of the SFRs has a diameter greater than $20''$, whereas in new galaxies, the diameters of the SFRs are usually significantly less than this value. Note that both in the center and on the periphery of the galaxy NGC 628, we have selected regions of the same size on average, except for some excess of small regions in the outer part of the galaxy, which may be due to selection effects.

The histogram in Fig.~\ref{aperdiam}a is more representative of the process of isolating SFRs than of their physical characteristics. Figure ~\ref{aperdiam}b shows the distribution of SFRs over linear diameters calculated for the distance estimates above, and here we see a slightly different picture. The largest SFRs ($\gtrsim 1$ kpc) are observed in the galaxies NGC 628 and NGC 660 (both in the disk and in the ring), while in the interacting galaxies the size of the SFRs is significantly smaller ($\lesssim 700$ pc). What was said above about the angular sizes of the central and peripheral SFRs in the galaxy NGC 628, is obviously true for their linear sizes. It should be clarified that the almost complete absence of identified SFRs in NGC 628, which are comparable in size to the SFRs in the galaxies NGC 4395 and NGC 4618, may be a selection effect. At a distance of NGC 628, a SFR diameter of about 250 pc corresponds to an aperture diameter of $5''$. In the images we used, we could have missed such small objects. Therefore, we can only say that the average size of SFRs in NGC 628 exceeds the average size of SFRs in NGC 1512, NGC 4395 and NGC 4618.

\begin{figure*}
    \begin{subfloat}
    \centering{\includegraphics[width=0.4\linewidth]{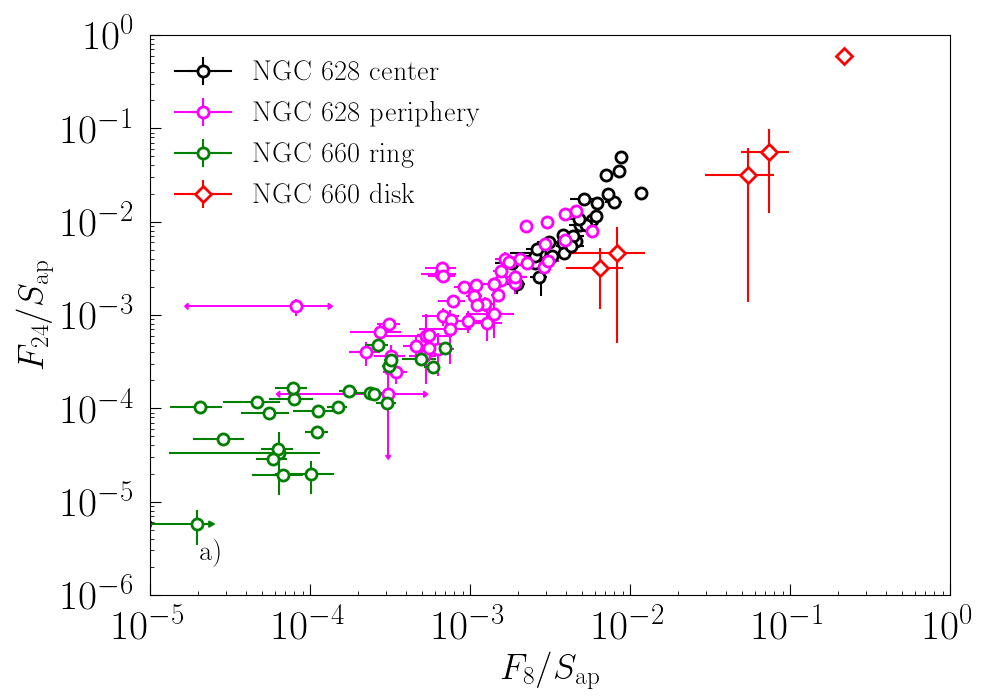}} 
    \end{subfloat}
    \begin{subfloat}
    \centering{\includegraphics[width=0.4\linewidth]{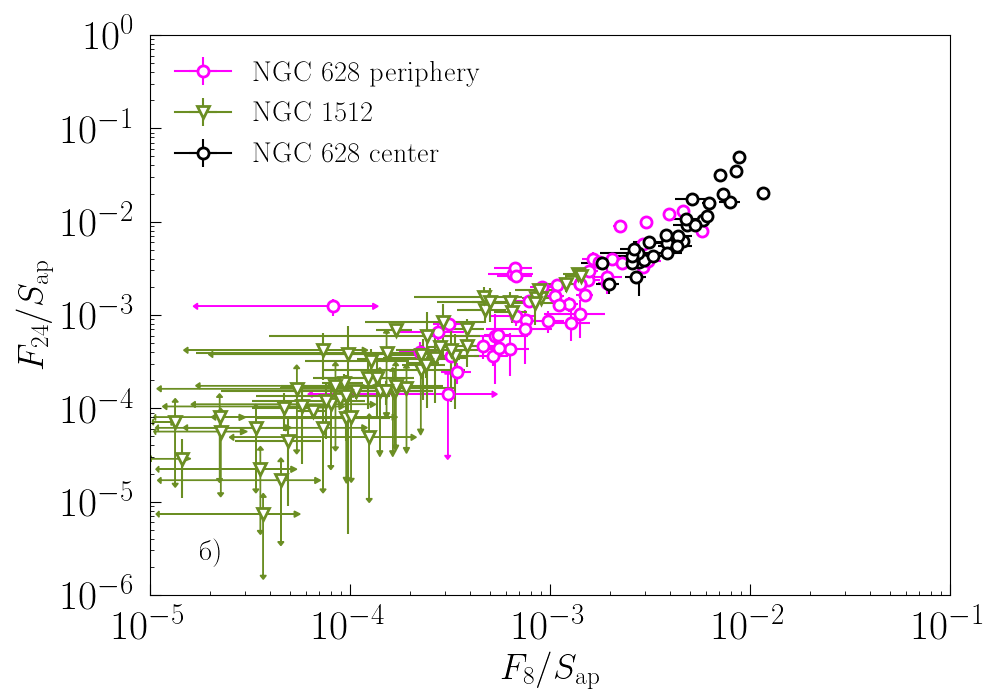}} 
    \end{subfloat}
    \begin{subfloat}
    \centering{\includegraphics[width=0.4\linewidth]{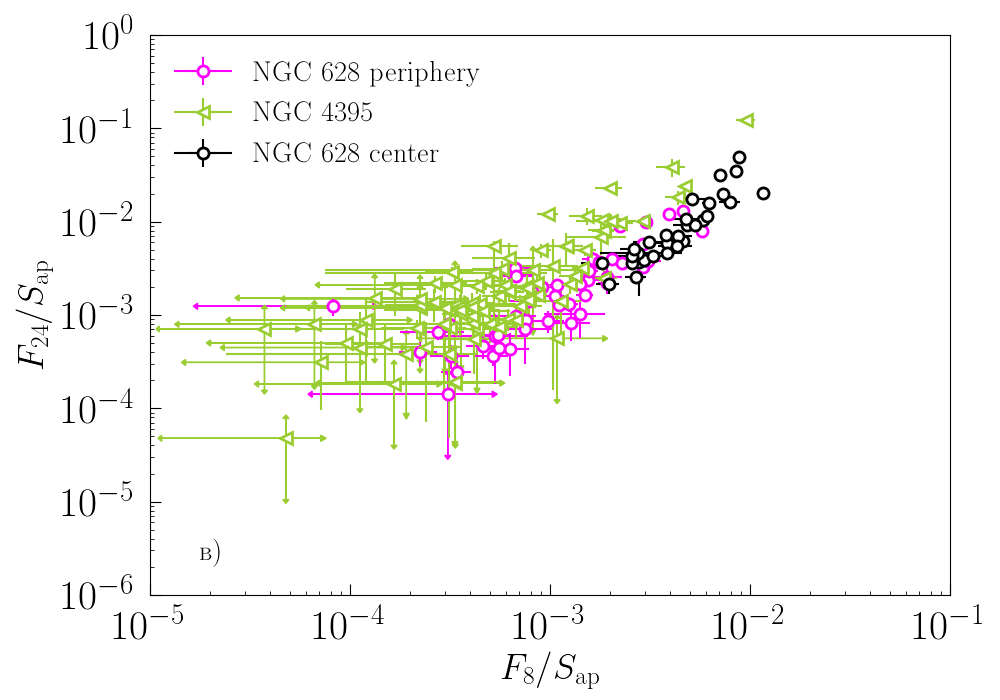}} 
    \end{subfloat}
        \begin{subfloat}
    \centering{\includegraphics[width=0.4\linewidth]{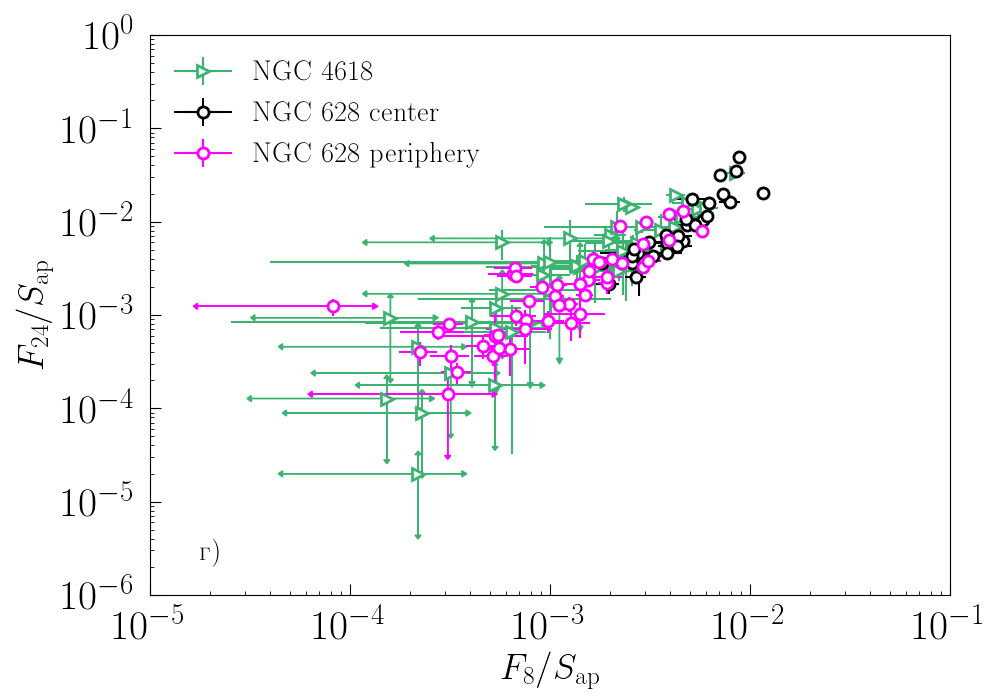}} 
    \end{subfloat}
\caption{The ratio of the surface brightness of radiation at wavelengths of 8 and 24 $\mu$m for SFRs from the galaxies studied in this paper.}
\label{F8_24_surfbri}
\end{figure*}

Another parameter that can be used to compare the systems in question is the surface brightness. Figure~\ref{F8_24_surfbri} compares the surface brightness (in Jy/pc$^2$)) of the studied SFRs at wavelengths of 8 and 24 $\mu$m separately for each galaxy studied in this work. The surface brightness of three SFRs of the NGC 660 galaxy disk at a wavelength of 8 $\mu$m exceeds the surface brightness of all other considered SFRs, but this may be due to the uncertain isolation of SFRs in this subsystem. Two more SFRs of the NGC 660 disk do not differ in surface brightness from the SFRs in the comparison galgalaxy. In contrast, the SFRs of the NGC 660 ring have significantly lower surface brightness at wavelengths of 8 and 24 $\mu$m, less than $10^{-3}$ Jy/pc$^2$, than the SFRs of the center and periphery of NGC628. The same low values of $F_{8}/S_{\rm ap}$ and $F_{24}/S_{\rm ap}$ are observed in the galaxy NGC 1512. In contrast to these two systems, the SFRs in the galaxies NGC 4395 and NGC 4618 practically do not differ in surface brightness from the SFRs from the comparison galaxy.

\begin{figure*}
\begin{subfloat}
\centering{\includegraphics[width=0.4\linewidth]{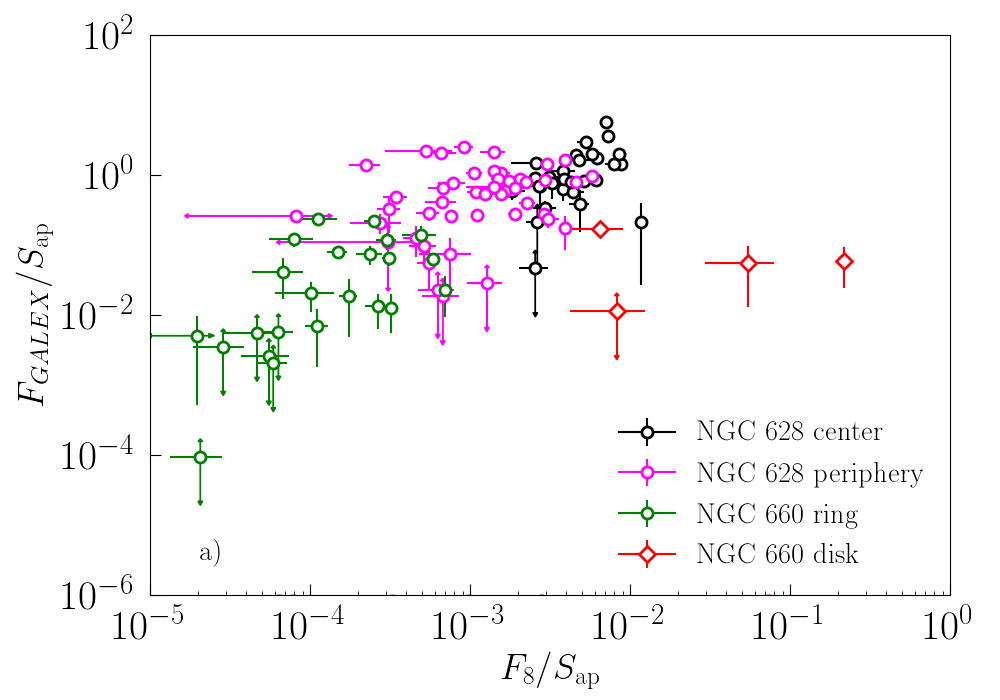}}
\end{subfloat}
\begin{subfloat}
\centering{\includegraphics[width=0.4\linewidth]{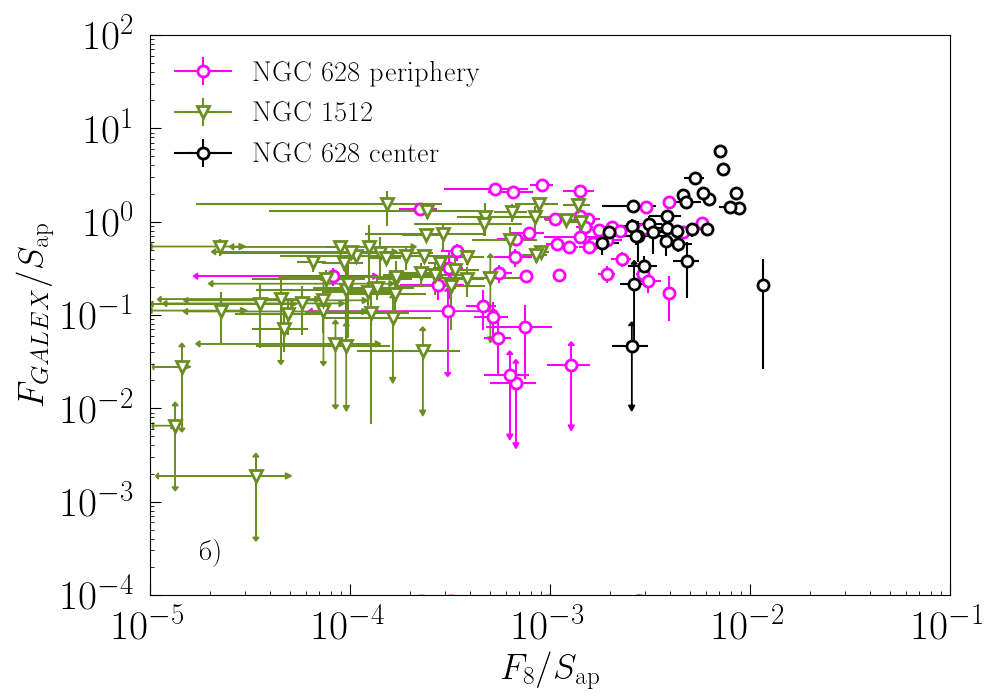}} 
\end{subfloat}
\begin{subfloat}
\centering{\includegraphics[width=0.4\linewidth]{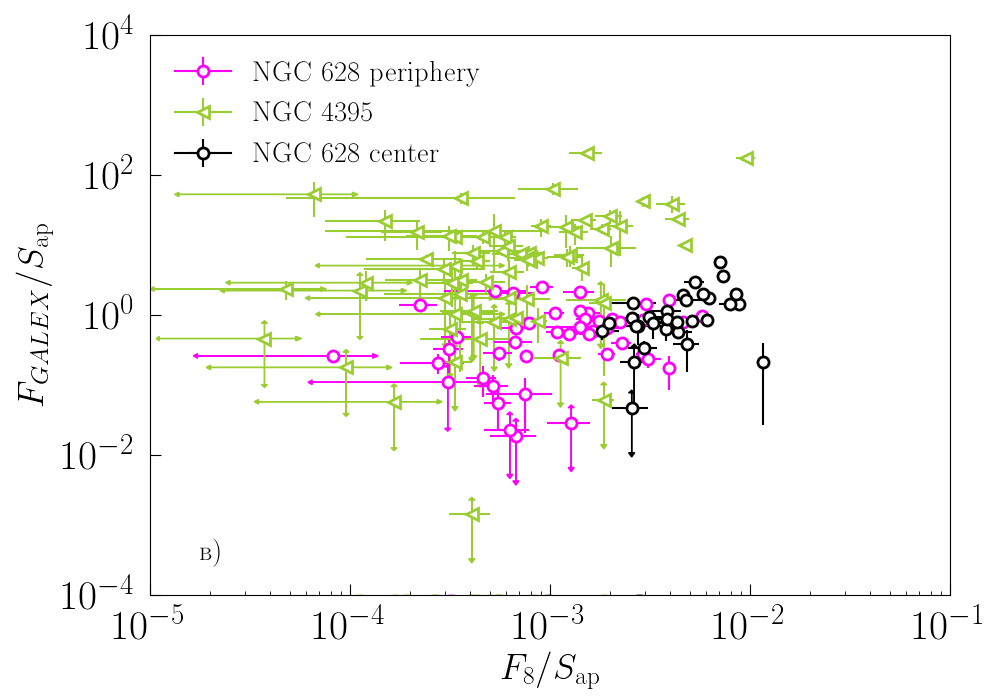}} 
\end{subfloat}
\begin{subfloat}
\centering{\includegraphics[width=0.4\linewidth]{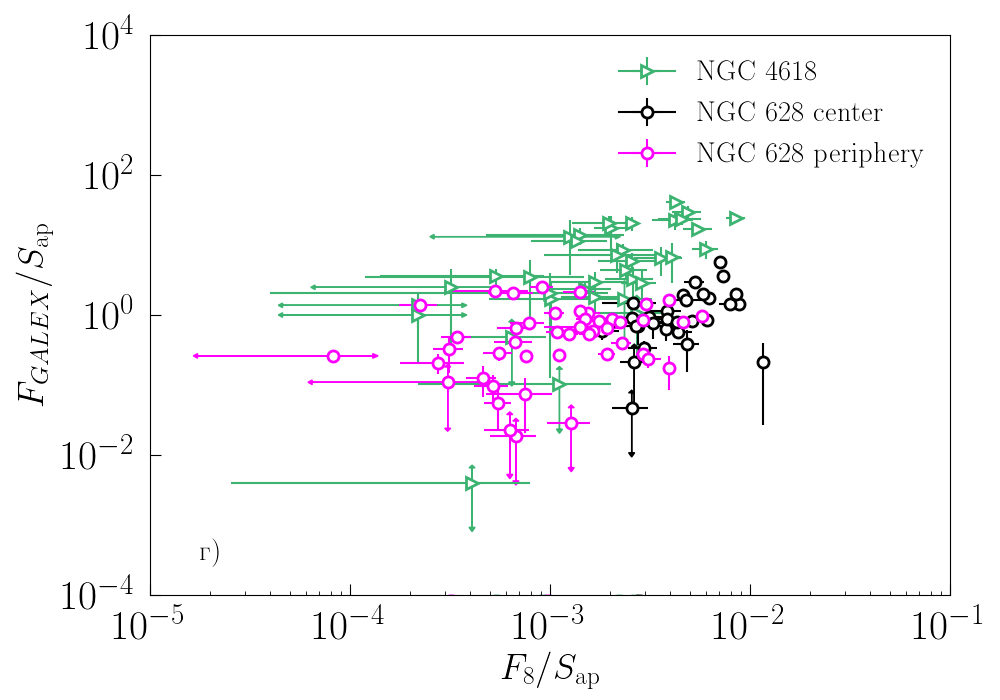}}
\end{subfloat}
\caption{The ratio of surface brightness at a wavelength of 8 $\mu$m and in the UV range for SFRs in the galaxies studied in this paper.}
\label{F8_24_GALEX}
\end{figure*}

The low surface brightness of IR radiation of the SFRs in NGC 1512 and in the NGC 660 ring may be due to both a small number of sources (aromatic particles) and a low intensity of ultraviolet radiation. To compare the radiation in the infrared and ultraviolet ranges, in Fig.~\ref{F8_24_GALEX} we compare the results of aperture photometry of data from the archives of observations with the Spitzer and GALEX telescopes. The data of ultraviolet observations in the FUV filter and radiation at a wavelength of 8 $\mu$m are presented (for fluxes in the NUV bands and at a wavelength of 24 $\mu$m, the results are essentially the same).

As in Fig.~\ref{F8_24_surfbri}, we see a significant difference between the galaxies NGC 660 and NGC 1512 on the one hand and the galaxies NGC 4395 and NGC 4618 on the other hand. It is obvious that the intensity of the aromatic bands in the NGC 660 ring and in NGC 1512 correlates well with the UV emission in the filter FUV. In the NGC 660 ring, the UV flux from SFRs is significantly lower than in the comparison galaxy. The SFRs in NGC 1512 overlap somewhat with the SFRs in NGC 628, but on average the surface brightness of SFRs in this galaxy in both IR and UV is lower than in NGC 628.

\begin{figure}
\includegraphics[width=0.8\linewidth]{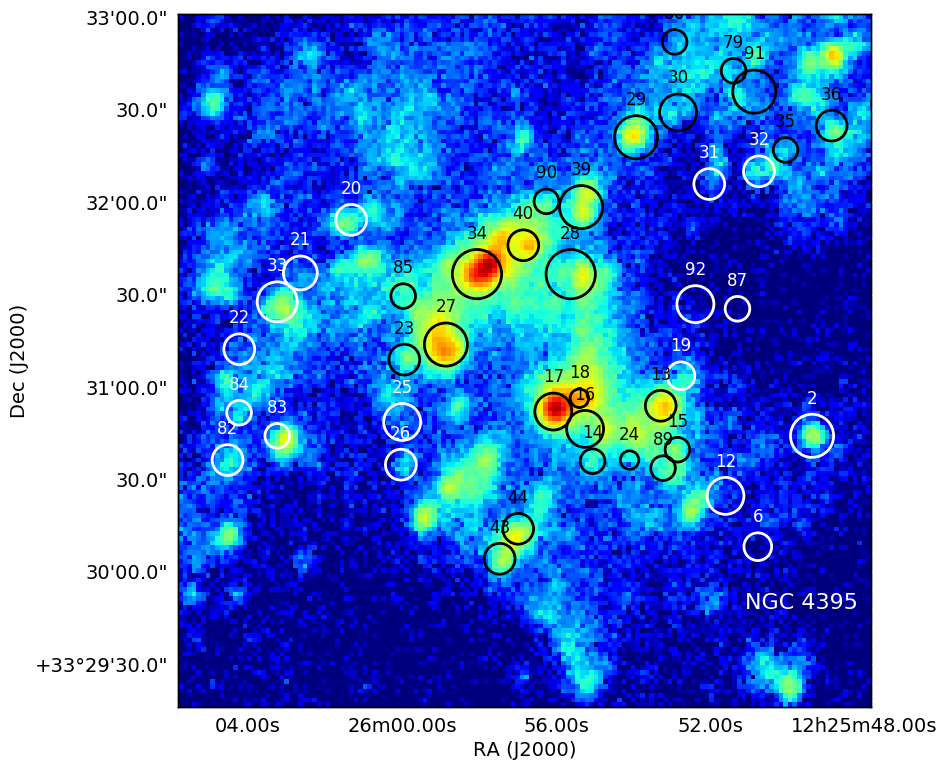}
\caption{A fragment of the image of the galaxy NGC 4395 in the UV range (GALEX FUV) with the studied SFRs.}
\label{NGC4395_NGC1512_GALEX}
\end{figure}

There is a temptation to link the observed surface brightness in the IR and UV ranges, but we must take into account that, firstly, in the galaxies NGC 4395 and NGC 4618, we observe a different picture, namely: a significant shift of the SFR positions in Fig.~\ref{F8_24_GALEX} up along the Y axis, which means a higher surface brightness in the FUV filter at the same surface brightness in the band 8 $\mu$m. Secondly, since the UV range was not taken into account when isolating SFRs, there is no noticeable UV radiation in some of the regions we have selected, as illustrated in Fig.~\ref{NGC4395_NGC1512_GALEX}, where the positions of SFRs in the NGC 4395 galaxy are marked in its images in the UV range. It can be seen that the UV radiation is not associated with some SFRs or is significantly shifted relative to them (for example, SFRs 6, 31, 87, 92).

\begin{figure*}
    \begin{subfloat}
    \centering{\includegraphics[width=0.4\linewidth]{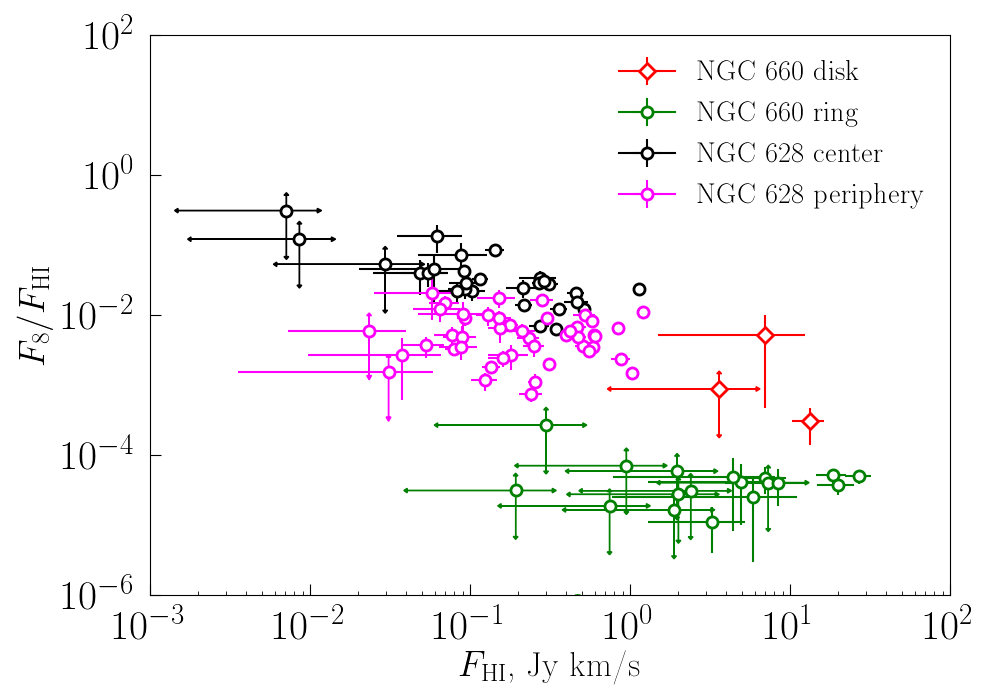}} 
    \end{subfloat}
    \begin{subfloat}
    \centering{\includegraphics[width=0.4\linewidth]{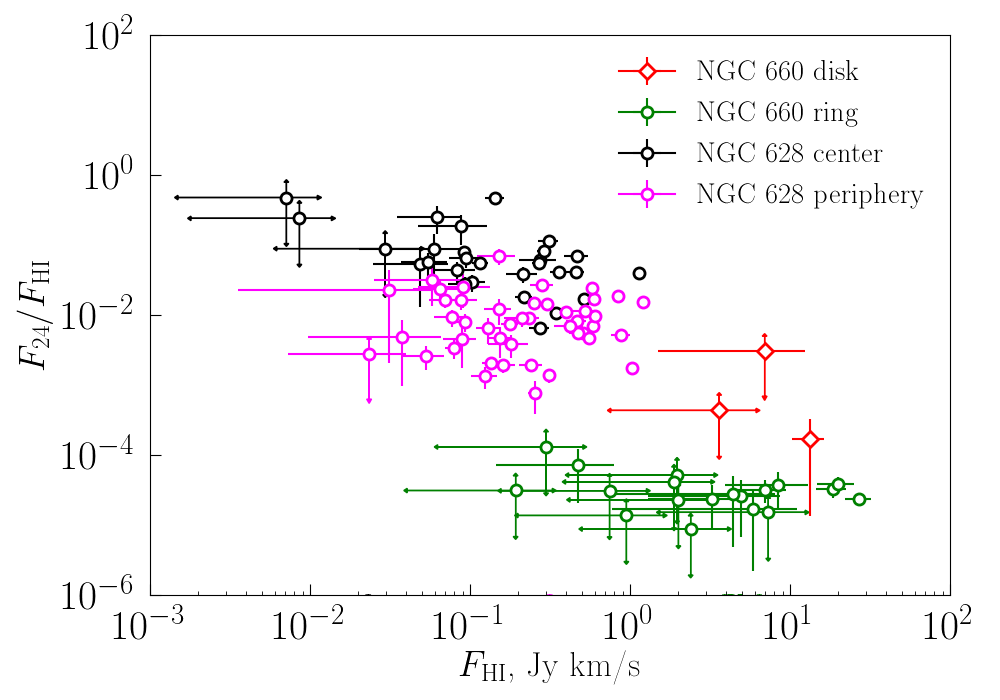}}
    \end{subfloat}
\caption{The ratio of fluxes $F_8/F_{\rm HI}$ and $F_{24}/F_{\rm HI}$ for SFRs in the galaxy NGC660 and the comparison galaxy as a function of the
flux in the neutral hydrogen line.}
\label{ngc660_HI}
\end{figure*}

Radiation in the near-IR, UV bands, and radiation in the H$\alpha$ line are associated with active star formation. Due to the availability of data on observations of neutral hydrogen in the galaxy NGC 660, we are able to correlate the emission in the IR range with the emission in the line at a wavelength of 21 cm, just as in the work \cite{Smirnova2017}, normalizing the flux in the IR range to the flux in the line HI. The results are shown in Fig.~\ref{ngc660_HI}.

Both in the diagram for a wavelength of 8 $\mu$m, and in the diagram for a wavelength of 22 $\mu$m, the SFRs of the NGC 660 ring as a group fall on the same (anti)correlation as the SFRs of the galaxies studied in \cite{Smirnova2017}: the brighter the region in the line HI, the lower its relative brightness in the IR range. \cite{Smirnova2017} suggested that the existence of such anticorrelation may be due to age in the sense that neutral hydrogen is particularly rich in young regions where conditions have not yet developed for generating intense infrared radiation. However, there is no correlation between $F_{8,22}/F_{\rm HI}$ and $F_{\rm HI}$ within these regions. Low fluxes in the IR range are observed both in SFRs rich in hydrogen and in SFRs poor in it.

\subsection{Kinematics}

In \cite{2019ARep...63..445S}, the value of the velocity scatter $\Delta V$, equal to the velocity difference of the extreme channels in the full spectrum, in which the signal exceeds the tripled standard deviation, was introduced as a characteristic of internal motions in the SFRs. The standard deviation was estimated for the part of the spectrum where there is no signal from the SFRs. Of the new galaxies, these data are only available for NGC 660. Also, only for this galaxy there are spectra in the  H$\alpha$  line. For them, the procedure described above for estimating $\Delta V$ does not work, and we used a more traditional approach, determining the width of the line by fitting the Gaussian into those spectra that are dominated by a single line. In this case, $\Delta V$ is estimated as FWHM for the fitted pro file, subtracting the width of the tool profile.

To estimate the contribution of the large-scale velocity field (beam smearing), we assumed that atomic hydrogen moves in the disk and ring in the same way as HII, i.e. it shows almost regular rotation.
The smoothed velocity field H$\alpha$ was used to analyze the distribution of velocities in pixels that fell into the elongated beam of WSRT observations, and the dispersion of these velocities in our apertures was calculated. As expected, in the outer regions of the ring, the correction is small and is 8–12 km/s. In the disk, its value is greater, 30–60 km/s, with maximum values near the intersections of the disk with the ring. In all cases, the estimated contribution of the large-scale velocity field is significantly less than the value $\Delta V$ in the same aperture, so we ignored it in the further analysis. Unfortunately, in the absence of statistical interpretation of the value $\Delta V$, the procedure for subtracting the contribution of the large-scale velocity field from it is unclear, which makes our conclusions somewhat uncertain.

\begin{figure}[t!]
    \begin{subfloat}
    \centering{\includegraphics[width=0.8\linewidth]{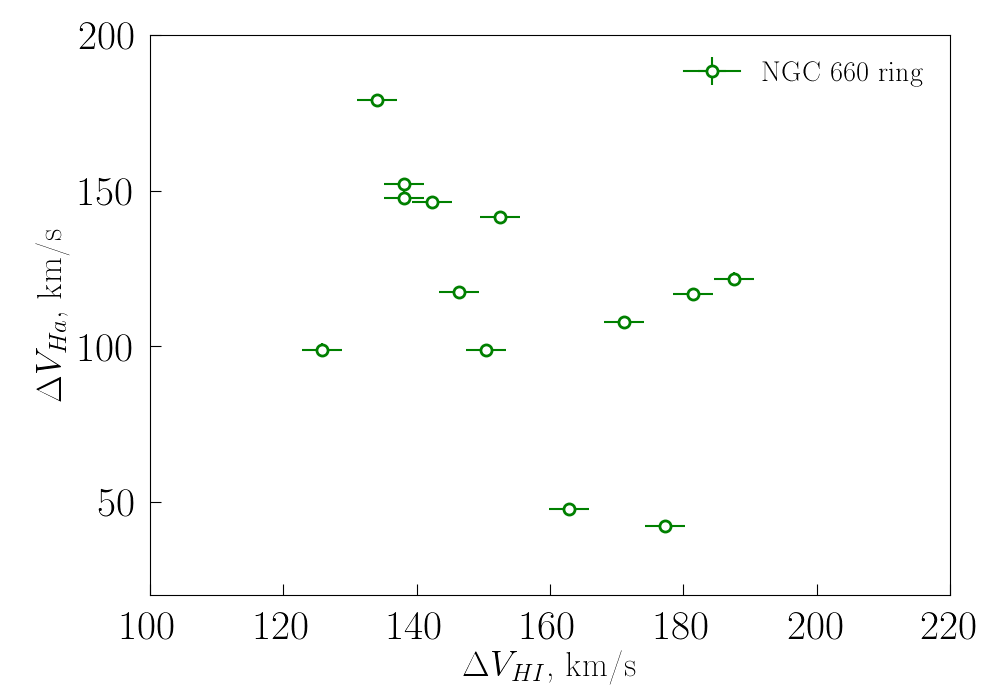}} 
    \end{subfloat} 
\caption{The ratio of velocity scatters in the lines H$\alpha$ and HI.}
\label{NGC660kincompar}
\end{figure}

\begin{figure}
    \begin{subfloat}
    \centering{\includegraphics[width=0.8\linewidth]{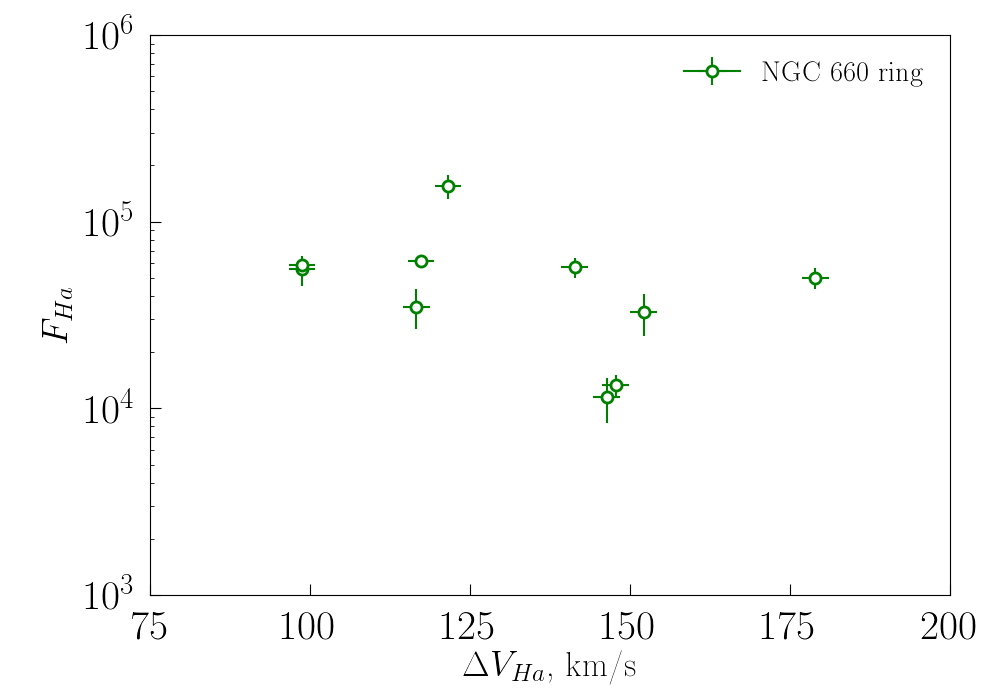}} 
%    \centering{\includegraphics[width=0.9\linewidth]{dvf.eps}} 
    \end{subfloat}
   \begin{subfloat}
    \centering{\includegraphics[width=0.8\linewidth]{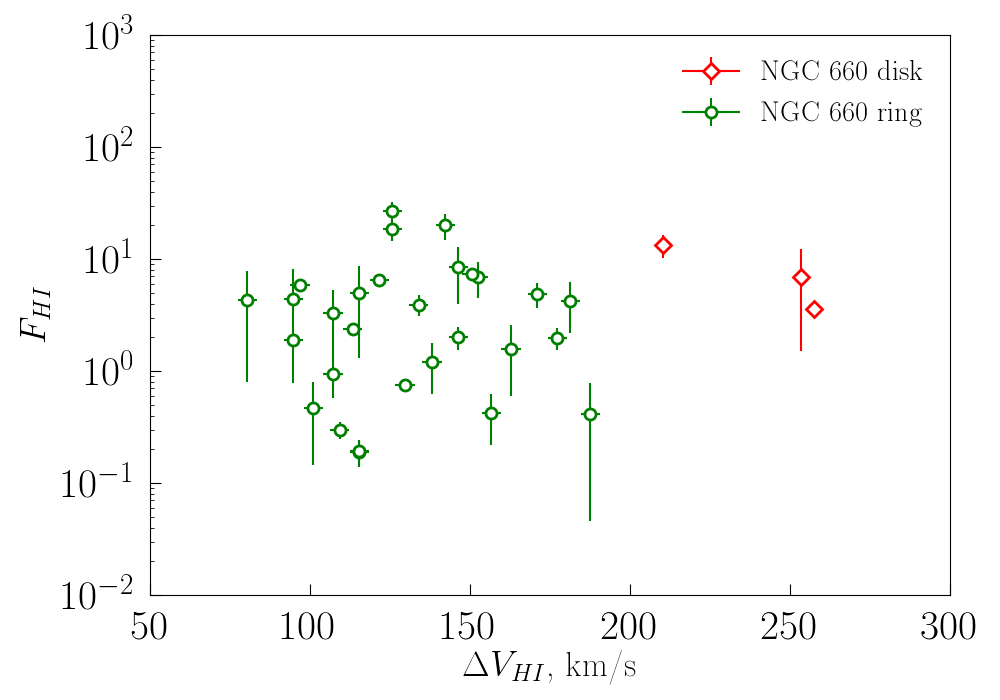}}
    \end{subfloat}
\caption{The ratio between the velocity scatter and the radiation flux in the SFRs from the galaxy NGC 660: HI (a) and H$\alpha$ (b).}
\label{NGC660kinematique}
\end{figure}

The values $\Delta V$ for H$\alpha$ and HI for NGC 660 are compared in Fig.~\ref{NGC660kincompar} and compared to the corresponding fluxes in Fig.~\ref{NGC660kinematique}.

%    10    fig

Figure~\ref{NGC660kincompar} shows that the velocity scatters in the SFRs of the NGC 660 ring in both lines are enclosed in the range from 100 to 200 km/s, but there is no correlation between them. The only region where the velocity scatter in the H$\alpha$ line exceeds 400 km/s, is characterized by a two-peak profile. Given its large size, this may mean that two regions fell into the aperture (interestingly, the velocity scatter in the line HI has a moderate value).

The ratio between flux and velocity scatter in the H$\alpha$  spectrum (Fig.~\ref{NGC660kinematique}a) shows a known correlation: the flux growth is accompanied by an increase in velocity scatter, which may indicate a relationship between velocity scatter and mass, although the nature of this correlation remains a subject of discussion \cite{2015MNRAS.449.3568M}. But there is almost no correlation between the velocity scatter in the lines HI and the corresponding flux (Fig.~\ref{NGC660kinematique}b).

\section{DISCUSSION}

Initially, the motivation for this work was the study of the polar-ring galaxy NGC660, which revealed a significant difference in the properties of infrared radiation in the SFRs of the disk and the ring of this galaxy. The assumption that this difference is due to the different nature of star formation in the two subsystems of NGC 660, was the incentive for this study, in which we expected to find similar differences in other galaxies that have recently experienced an episode of enhanced star formation due to some external influence. However, the real picture was more complicated.

A direct comparison of IR fluxes does show that in all new galaxies the SFRs are dimmer than the SFRs in the comparison galaxy. However, the reasons for this are different. The SFRs in the NGC 660 ring are
comparable in physical size to the SFRs in the “normal” galaxy NGC 628 and to the SFRs in the NGC 660 disk and their small IR flux is associated with low surface brightness. The SFRs in the galaxies NGC 4395 and NGC 4618 are similar in surface brightness to the SFRs in the NGC 628 galaxy, and their low fluxes are associated with smaller physical sizes. Finally, SFRs in the galaxy NGC 1512 are characterized by both small size and low surface brightness. These differences must obviously be related to the evolutionary differences between the galaxies in question. Let us look at them in more detail.

The galaxy NGC 1512 is a system with two rings—external and internal,—the morphology of which is largely determined by the interaction with the nearby dwarf galaxy NGC 1510. The distribution of gas and dust in the NGC 1512/NGC 1510 system was considered in \cite{2009MNRAS.400.1749K} and \cite{2010AJ....139.1190L}, respectively. \cite{2009MNRAS.400.1749K} showed that atomic hydrogen extends to a distance at least four times the optical size of the galaxy ($B_{25}$), and is largely concentrated in two spiral arms (Arm1 and Arm2). Individual clouds of HI are traced to distances of the order of $\sim30^\prime$ (for the adopted distance, this corresponds to 100 kpc). Up to about the same distance, UV radiation spots are also visible, probably tracking the position of star-forming regions (their more detailed analysis is presented in \cite{2015MNRAS.450.3381L}). At the same time, \cite{2009MNRAS.400.1749K} note that in the outer region of the NGC 1512 disk the radiation at a wavelength of 8 $\mu$m is not detected, which is consistent with our results: according to radiation maps at a wavelength of 8 $\mu$m, we were not able to isolate any SFRs in the outer region of this galaxy. In \cite{2010AJ....139.1190L}, the outer regions of NGC 1512 were also not considered. Thus, there is a similarity between the galaxies NGC 660 and NGC 1512: the SFRs in the rings of NGC 660 and in NGC 1512 are characterized by reduced surface brightness at a wavelength of 8 $\mu$m. Moreover, on the
periphery of NGC 1512, exposed to the tidal action of NGC 1510, the surface brightness at a wavelength of 8 $\mu$m is so low that radiation is not detected at all, although the content of HI there, as in the NGC 660 ring (see Fig.~\ref{ngc660_HI}, is high. Thus, the common characteristic of the galaxies NGC 660 and NGC 1512 can be considered the presence of subsystems (the NGC 660 ring, external spiral arms of NGC 1512), characterized by a significant content of atomic hydrogen, the presence of UV radiation sources and low surface brightness at a wavelength of 8 $\mu$m.

It should be noted that the tidal structures in NGC 1512, as well as the NGC 660 ring, are not longlived structures, whereas in lopsided galaxies the characteristic time of star formation initiation may be longer. However, the problem of the age of the considered SFRs requires a separate study.

A common feature of the galaxies NGC 4395 and NGC 4618 is their lopsidedness. There is no clear answer to the question about the nature of the asymmetric structure of disk galaxies. Numerical calculations
show that in some cases such asymmetry may occur during the dynamic evolution of the disk in barred spiral galaxies.
Among the more common explanations are tidal interactions with other galaxies or an asymmetric inflow of external matter \cite{2009PhR...471...75J}. One is tempted to associate the above-mentioned features of NGC 4395 and NGC 4618 with their asymmetry. However, they differ from other galaxies considered in this article by at least three other parameters: distance, mass, and star formation rate, and the differences in mass and star formation rate may be related.

The close location of NGC 4395 and NGC 4618 could lead to the fact that in these galaxies, due to the better spatial resolution, we isolated smaller SFRs. Indeed, in the galaxy NGC 4395, it is possible to specify several groups of SFRs, which at a lower resolution would be identified by us as one larger SFR. However, there are few such groups, and they would not change the overall picture of the distribution of SFRs by size. In addition, the relatively small physical size is also characteristic of SFRs in the galaxy NGC 1512, located much further away. We have already noted the possible effects of selection when isolating regions in NGC 628, but even with their consideration, the main conclusion remains the same: the average size of SFRs in NGC 4395 and NGC 4618 is inferior to the average size of SFRs in other galaxies.

Both of the above reasons of galaxy asymmetry can cause increased star formation, and we do see some indications of this. The total luminosity of NGC 4395 and NGC 4618 in the mid-IR range indicates that the integrated star formation rate in them is an order of magnitude or more inferior to the star formation rate in the galaxies NGC 628 and NGC 1512 (see Table~\ref{distances}), however, this difference indicates a smaller size of NGC 4395 and NGC 4618: the total surface brightness of these galaxies at a wavelength of 24 $\mu$m (according to our estimates) and in the B band (according to the HyperLEDA database) exceeds the similar parameters of the galaxies NGC 628 and NGC C1512. In other words, in the two lopsided galaxies considered in this paper, the star formation is actually somewhat more intense than in the other galaxies. In addition, the SFRs in NGC 4395 and NGC 4618 have a higher surface brightness in the UV range than the SFRs in the comparison galaxy, which have the same brightness in the near IR range. Thus, although the asymmetry of disk galaxies is a fairly common property (\cite{2009PhR...471...75J}), in two specific cases (NGC 4395 and NGC 4618) we see noticeable differences from both interacting and “normal” galaxies. It should be noted that the galaxy NGC 4395 is isolated, and the galaxy NGC 4618 although usually considered in conjunction with the galaxy NGC 4625, may also not be interacting (\cite{2012AJ....144...67K}). It is obvious that the interaction is not the only reason for the formation of SFRs with a small flux in the IR range.

It should be noted that, if in the galaxy NGC 628 we separately consider SFRs in the center and on the periphery, the above diagrams also show differences between them: central SFRs exhibit higher luminosities and surface brightnesses in the near IR region than peripheral SFRs, however, the differences noted above between the galaxies under consideration and the galaxy NGC 628 turn out to be significantly larger than the differences between the central and peripheral SFRs in NGC 628. In the UV range, as well as in the emission of neutral hydrogen, there are no systematic differences between the central and peripheral SFRs in NGC 628.

The metallicities of all the galaxies under consideration are close, so we cannot expect any significant differences related to the chemical composition. The galaxy NGC 660 stands somewhat apart in this respect. Its metallicity (as well as other integral characteristics) refers mainly to the disk. We have not found any determinations of the chemical composition of the ring in the literature, except for the work \cite{2004A&A...421..833K}, the authors of which showed that the stellar population of the ring is best described by isochrones with a rather low metallicity, $Z=0.008$. In addition, \cite{2000A&A...357..443A} estimated the ratio of dust and gas masses in the NGC 660 ring. It turned out that its value is 2 to 3 times lower than in the solar neighborhood. If we assume that this ratio is a measure of metallicity, it also indicates the content of heavy elements just a few times lower than the solar ($12+\lg({\rm O}/{\rm H})\sim8.1\div8.3$) Thanks to the observations in HI presented in the press release\footnote{https://www.astron.nl/dailyimage/main.php?date=20100923} \cite{Jozsa2010} it is clear that the ring in NGC 660 was formed as a result of gas capture from the disk of a rather massive dwarf irregular galaxy UGC 1195. Direct measurements of the metallicity of this galaxy are not available in the literature, but the using of the “metallicity luminosity” relation from \cite{2004A&A...425..849P} gives $M_B({\rm UGC~1195)}=-17.67$ for (according to the HyperLeda database) the value $12+\lg({\rm O}/{\rm H})=8.33$ in good agreement with the above estimate of metallicity of the NGC 660 polar ring. At the same time, as our research also shows \cite{Smirnova2017}, differences that are related to the metallicity become significant at values of $12+\lg({\rm O}/{\rm H})<8$.

\section{CONCLUSIONS}

The paper considers the parameters of SFRs in several galaxies and possible relationships between these parameters. Based on the results obtained, the following conclusions are made:
\begin{enumerate}
\item{In the galaxies with signs of recent interaction NGC 660 and NGC 1512, the surface brightness of star-forming complexes in the UV and IR ranges is significantly lower than in the comparison galaxy. At the same time, in the peripheral region of NGC 1512, there are no SFRs emitting in the near-IR range at all.}
\item{In the asymmetric galaxies NGC 4395 and NGC 4618, the surface brightness of SFRs in the IR range does not differ from the surface brightness of SFRs in “normal” galaxies. The surface brightness of SFRs in these galaxies in the UV range exceeds the similar brightness of SFRs in comparison galaxies. However, the physical size of the SFRs in these two galaxies is smaller than the size of the SFRs in both the interacting galaxies and the comparison galaxy.}
\item{In the SFRs of the ring of the NGC 660 galaxy, the HI flux at 21 cm is comparable to or exceeds the flux in the SFRs from the comparison galaxy (at significantly lower IR fluxes). The velocity scatter in the line H$\alpha$  increases with the flux in this line; the velocity scatter in the line HI does not depend on the flux in it.}
\end{enumerate}

The results of the study may indicate that the parameters of star-forming complexes differ in galaxies with signs of recent interaction, in galaxies with asymmetric disks and in normal galaxies. The differences observed in both the physical dimensions and the surface brightness of the SFRs indicate that the course of star formation process depends on many factors, and not only on the belonging of the galaxy to one or another morphological type.

The authors thank the referee for remarks and comments that allowed a deeper understanding of the results presented, as well as A. Alakoz and D. Makarov for useful discussions. This work is based in part on observations made with the Spitzer Space Telescope, which was operated by the Jet Propulsion Laboratory, California Institute of Technology under a contract with NASA and Herschel Space Observatory that is an ESA space observatory with science instruments provided by European-led Principal Investigator consortia and with important participation from NASA. The astronomical database HYPERLEDA
(http://leda.univ-lyon1.fr) was used in the work.

The research was supported by Russian Foundation for Basic Research as part of the science project no.~19-32-50063. The work is based on the use of observations from 6-meter telescope of SAO RAS, supported by the Ministry of Science and Higher Education of the Russian Federation (agreement no.~05.619.21.0016, project ID RFMEFI61919X0016).

\clearpage

\end{document}